\def\year{2022}\relax
\documentclass[letterpaper]{article} 
\usepackage{aaai22}  
\usepackage{times}  
\usepackage{helvet}  
\usepackage{courier}  
\usepackage[hyphens]{url}  
\usepackage{graphicx} 
\usepackage{natbib}  
\usepackage{caption} 
\DeclareCaptionStyle{ruled}{labelfont=normalfont,labelsep=colon,strut=off} 
\usepackage{algorithm}
\usepackage{algorithmic}
\usepackage[utf8]{inputenc} 
\usepackage[T1]{fontenc}    
\usepackage{url}            
\usepackage{booktabs}       
\usepackage{amsfonts}       
\usepackage{nicefrac}       
\usepackage{microtype}      
\usepackage{xcolor}         
\usepackage{enumitem}
\usepackage{multirow}
\usepackage{graphicx}
\usepackage{amsmath}

\usepackage{tabularx}
\usepackage{newfloat}
\usepackage{listings}
\floatstyle{ruled}
\newfloat{listing}{tb}{lst}{}
\floatname{listing}{Listing}
\usepackage{arydshln}
\usepackage{enumitem}

\title{Improving Biomedical Information Retrieval with Neural Retrievers}
\author {
    Man Luo,\textsuperscript{\rm 1}
    Arindam Mitra,\textsuperscript{\rm 2}
    Tejas Gokhale,\textsuperscript{\rm 1}
    Chitta Baral\textsuperscript{\rm 1}
}
\affiliations {
    \textsuperscript{\rm 1} Arizona State University, 
    \textsuperscript{\rm 2} Microsoft\\
    \texttt{mluo26@asu.edu, arindam.mitra2@gmail.com, tgokhale@asu.edu, chitta@asu.edu}
}

\begin{document}

\maketitle

\begin{abstract}
Information retrieval (IR) is essential in search engines and dialogue systems as well as natural language processing tasks such as open-domain question answering. 
IR serve an important function in the biomedical domain, where content and sources of scientific knowledge may evolve rapidly.
Although neural retrievers have surpassed traditional IR approaches such as TF-IDF and BM25 in standard open-domain question answering tasks, they are still found lacking in the biomedical domain.
In this paper, we seek to improve information retrieval (IR) using neural retrievers (NR) in the biomedical domain, and achieve this goal using a three-pronged approach.
First, to tackle the relative lack of data in the biomedical domain, we propose a template-based question generation method that can be leveraged to train neural retriever models.
Second, we develop two novel pre-training tasks that are closely aligned to the downstream task of information retrieval.
Third, we introduce the ``Poly-DPR'' model which encodes each context into multiple context vectors.
Extensive experiments and analysis on the BioASQ challenge suggest that our proposed method leads to large gains over existing neural approaches and beats BM25 in the small-corpus setting.
We show that BM25 and our method can complement each other, and a simple hybrid model leads to further gains in the large corpus setting.
\end{abstract}

\section{Introduction}
\begin{table*}[t]
    \centering 
    \small
    \resizebox{\linewidth}{!}{
    \begin{tabular}{@{}p{0.175\linewidth} p{0.32\linewidth} p{0.33\linewidth} p{0.275\linewidth}@{}}
        \toprule
        \textbf{Question} & \textbf{Answer} & \textbf{Retrieved Context (BM25)} & \textbf{Retrieved Context(DPR)} \\
        \toprule
    What is Soluvia? 
        & Soluvia by Becton Dickinson is a microinjection system for intradermal delivery of vaccines.
        & The US FDA approved Sanofi Pasteur's Fluzone Intradermal influenza vaccine that uses a new microinjection system for intradermal delivery of vaccines (Soluvia, Becton Dickinson).
        & Internet-ordered viagra (sildenafil citrate) is rarely genuine. \\
    \midrule
    Is BNN20 involved in Parkinson's disease?
        & BNN-20 could be proposed for treatment of PD
        & Rare causes of dystonia parkinsonism.
        & BNN-20 could be proposed for treatment of PD \\
    \midrule
    How large is a lncRNAs?
        & lncRNAs are defined as RNA transcripts longer than 200 nucleotides that are not transcribed into proteins.
        & lncRNAs are closely related with the occurrence and development of some diseases.
        & An increasing number of long noncoding RNAs (lncRNAs) have been identified recently. \\
    \bottomrule
    \end{tabular}
    }
    \caption{Illustrative examples from the BioASQ challenge along with the context retrieved by two methods BM25 and DPR.
    }
    \label{tab:examples_bioasq}
\end{table*}

Information retrieval (IR) is widely used in commercial search engines and is an active area of research for natural language processing tasks such as open-domain question answering (ODQA).
IR has also become important in the biomedical domain due to the explosion of information available in electronic form~\citep{shortliffe2014biomedical}.
Biomedical IR has traditionally relied upon term-matching algorithms (such as TF-IDF and BM25~\citep {Robertson2009ThePR}), which search for documents that contain terms mentioned in the query.
For instance, the first example in Table~\ref{tab:examples_bioasq} shows that BM25 retrieves a sentence that contains the word \textit{``Soluvia''} from the question.
However, term-matching suffers from failure modes, especially for terms which have different meanings in different contexts (example 2), or when crucial semantics from the question are not considered during retrieval (for instance, in the third example when the term ``how large'' is not reflected in the answer retrieved by BM25).

Since these failure modes can have a direct impact on downstream NLP tasks such as open-domain question answering (ODQA), there has been interest in developing neural retrievers (NR)~\citep{Karpukhin2020DensePR}.
NRs which represent query and context as vectors and utilize similarity scores for retrieval, have led to state-of-the-art performance on ODQA benchmarks such as Natural Questions~\citep{Kwiatkowski2019NaturalQA} and TriviaQA~\citep{Joshi2017TriviaQAAL}. 
Unfortunately, these improvements on standard NLP datasets are not observed in the biomedical domain with neural retrievers.

Recent work provides useful insights to understand a few shortcomings of NRs.
\citet{Thakur2021BEIRAH} find NRs to be lacking at exact word matching, which affects performance in datasets such as BioASQ~\citep{Tsatsaronis2015AnOO} where exact matches are highly correlated with the correct answer.
\citet{Lewis2021QuestionAA} find that in the Natural Questions dataset, answers for $63.6\%$ of the test data overlap with the training data and DPR performs much worse on the non-overlapped set than the test-train overlapped set. 
In this work, we found this overlap to be only $2\%$ in the BioASQ dataset, which could be a potential reason for lower performance of NR methods.
We also discovered that NRs produce better representations for short contexts that for long contexts -- when the long context is broken down into multiple shorter contexts, performance of NR models improves significantly.

In this paper, we seek to address these issues and improve the performance of neural retrieval beyond traditional methods for biomedical IR.
While existing systems have made advances by improving neural re-ranking of retrieved candidates~\citep{almeida2020bit,pappas2020aueb}, our focus is solely on the retrieval step, and therefore we compare our neural retriever with other retrieval methods.
Our method makes contributions to three aspects of the retrieval pipeline -- question generation, pre-training, and model architecture.

Our first contribution is the \textbf{``Poly-DPR''} model architecture for neural retrieval. 
Poly-DPR builds upon two recent developments: Poly-Encoder~\cite{Humeau2020PolyencodersAA} and Dense Passage Retriever~\citep{Karpukhin2020DensePR}.
In DPR, a question and a candidate context are encoded by two models separately into a contextual vector for each, and a score for each context can be computed using vector similarity.
On the other hand, Poly-Encoder represents the query by $K$ vectors and produces context-specific vectors for each query.
Instead, our approach Poly-DPR represents each \textit{context} by $K$ vectors and produces \textit{query-specific vectors} for each context.
We further design a simple inference method that allows us to employ MIPS~\citep{Shrivastava2014AsymmetricL} during inference.

Next, we develop \textbf{``Temp-QG''}, a template-based question generation method which helps us in generating a large number of domain-relevant questions to mitigate the train-test overlap issue.
TempQG involves extraction of templates from in-domain questions, and using a sequence-to-sequence model~\cite{Sutskever2014SequenceTS} to generate questions conditioned on this template and a text passage. 

Finally, we design two new pre-training strategies: \textbf{``ETM''} and \textbf{``RSM''} that leverage our generated dataset to pre-train Poly-DPR.
These tasks are designed to mimic domain-specific aspects of IR for biomedical documents which contain titles and abstracts, as opposed to passage retrieval from web pages~\cite{Chang2020PretrainingTF}.
Our pre-training tasks are designed to be used for long contexts and short contexts. 
In both tasks, we utilize keywords in either query or context, such that the capacity of neural retrievers to match important keywords can be improved during training.

Armed with these three modules, we conduct a comprehensive study of document retrieval for biomedical texts in the BioASQ challenge.
Our analysis demonstrates the efficacy of each component of our approach.
Poly-DPR outperforms BM25 and previous neural retrievers for the BioASQ challenge, in the small-corpus setting.
A hybrid method, which is a simple combination of BM25 and NR predictions, leads to further improvements.
We perform a post-hoc error analysis to understand the failures of BM25 and our Poly-DPR model.
Our experiments and analysis reveal aspects of biomedical information retrieval that are not shared by generic open-domain retrieval tasks.
Findings and insights from this work could benefit future improvements in both term-based as well as neural-network based retrieval methods.

\section{Related Work}

\textbf{Neural Retrievers} aim to retrieve relevant context from a large corpus given a query.
NRs can be clubbed into two architectural families -- cross-attention models~\citep{Nogueira2019PassageRW,MacAvaney2019CEDRCE,Yang2019EndtoEndOQ}, and dual-encoder models which employ separate encoders to encode the query and context~\citep{Karpukhin2020DensePR,Chang2020PretrainingTF}.
The cross-attention model requires heavy computation and can not be directly used in a large corpus setting, while dual-models can allow pre-computation of context representations and the application of efficient search methods such as MIPS~\cite{Shrivastava2014AsymmetricL} during inference.
To take advantage of both models, Poly-Encoder~\citep{Humeau2020PolyencodersAA} uses $K$ representations for each query and an attention mechanism to get context-specific query representations.
ColBERT~\citep{Khattab2020ColBERTEA} extends the dual-encoder architecture by performing a token-level interaction step over the query and context representations, but requires significant memory for large corpora~\citep{Thakur2021BEIRAH}.

\textbf{Pre-training Tasks for NR.} 
Masked language modeling (MLM) and next-sentence prediction introduced in BERT~\citep{Devlin2019BERTPO} have led to a paradigm shift in the training of neural network models for multiple NLP tasks.
For text retrieval, pre-training tasks that are more aligned with the retrieval task have been developed.
\citet{Chang2020PretrainingTF} propose Body First Selection (BFS), and Wiki Link Prediction (WLP) for document retrieval. 
\citet{Lee2019LatentRF} propose an Inverse Cloze Task (ICT) task in which a random sentence drawn from a passage acts as a query and the remaining passage as a relevant answer.
\citet{Guu2020REALMRL} show that ICT effectively avoids the cold-start problem.

\textbf{Question Generation (QG)} methods have become sophisticated due to the advances in sequence-to-sequence modeling~\citep{Sutskever2014SequenceTS}; QG is considered an auxiliary pre-training task for question answering models~\citep{Alberti2019SyntheticQC}. 
One set of QG methods can be categorized as `Answer-Aware' QG~\citep{Du2018HarvestingPQ,Zhao2018ParagraphlevelNQ,Dong2019UnifiedLM}, in which an answer extraction model first produces potential answers, followed by a question generator which generates a question given the context and a potential answer.
\citet{Alberti2019SyntheticQC} utilizes cycle consistency to verify whether a question-answering model predicts the same answer to the generated question.
A second set of QG methods generate questions without conditioning the generator using the answer --
for instance, \citet{Lopez2020TransformerbasedEQ} propose end-to-end question generation based on the GPT-2 model, while~\citet{lewis2019unsupervised,Fabbri2020TemplateBasedQG,banerjee2021self} generate questions using linguistic and semantic templates. 
Question paraphrasing~\cite{hosking-lapata-2021-factorising} is a related approach
for creating augmented training samples.
Question generation has also been explored in visual question answering, with end-to-end methods~\cite{li2018visual,krishna2019information} and template-based methods~\cite{banerjee2021weaqa}. 
While our proposed question generation method is also template-based, instead of using a pre-defined list of templates designed by humans, our template extraction process is automated.
\section{Poly-Dense Passage Retriever}
\label{sec:model}
\subsection{Preliminaries}
\paragraph{Dense Passage Representation} (DPR)~\citep{Karpukhin2020DensePR} is a neural retriever model belonging to the dual-model family.
DPR encodes the query $q$ and the context $c$ into dense vector representations:
\begin{equation}
    v_q = E_q(q)\texttt{[CLS]}, \quad v_c = E_c(c)\texttt{[CLS]}.
    \label{eq:dpr_enc}
\end{equation}
where $E_q$ and $E_c$ are BERT~\citep{Devlin2019BERTPO} models which output a list of dense vectors $(h_1, \dots, h_n)$ for each token of the input, and the final representation is the vector representation of special token$\texttt{[CLS]}$.
$E_q$ and $E_c$ are initialized identically and are updated independently while being trained with the objective of minimizing the negative log likelihood of a positive (relevant) context. 
A similarity score between $q$ and each context $c$ is calculated as the inner product between their vector representations:
\begin{equation}
    \mathrm{sim}(q, c) = v_q^T v_c.
    \label{eq:dpr_sim}
\end{equation}

\paragraph{Poly-Encoder}\citep{Humeau2020PolyencodersAA} also uses two encoders to encode query and context, but the query is represented by $K$ vectors instead of a single vector as in DPR.
Poly-Encoder assumes that the query is much longer than context, which is in contrast to information retrieval and open-domain QA tasks in the biomedical domain, where contexts are long documents and queries are short and specific. 

\subsection{Poly-DPR: Poly-Dense Passage Retriever}
\label{sec:k-code-dpr}
We integrate Poly-Encoder and DPR to use $K$ vectors to represent context rather than query.  
In particular, the context encoder includes $K$ global features $(m_1, m_2, \cdots, m_k)$, which are used to extract representation $v_c^i,~\forall i\in\{1 \cdots k\}$ by attending over all context tokens vectors. 
\begin{align}
    v_c^i &= \sum_{n} w_n^{m_i} h_n,~~\text{where}\\
    (w_1^{m_i} \dots, w_n^{m_i} ) &= \mathrm{softmax}(m_i^{T} \cdot h_1,  \dots, m_i^{T}\cdot h_n).
\end{align}

After extracting $K$ representations, a query-specific context representation  $v_{c,q}$ is computed by using the attention mechanism: 
\begin{align}
    v_{c,q} &= \sum_k w_k v_c^k,~~\text{where}\\
    (w_1, \dots, w_k) &= \mathrm{softmax}(v_q^{T}\cdot v_c^1, \dots, v_q^{T}\cdot v_c^k).
\end{align}

Although we can pre-compute $K$ representations for each context in the corpus, during inference, a ranking of the context needs to be computed after obtaining all query-specific context representations.
As such, we can not directly use efficient algorithms such as MIPS~\cite{Shrivastava2014AsymmetricL}. 
To address this challenge, we use an alternative similarity function for inference --
the score $\mathrm{sim_{infer}}$ is computed by obtaining $K$ similarity scores for the query and each of the $K$ representations, 
and take the maximum as the similarity score between context and query:
\begin{equation} 
    \mathrm{sim_{infer}}(q, c) = \mathrm{max} (v_q^T \cdot v_c^1, \dots, v_q^T \cdot v_c^k).
\label{eq:similarity_func}
\end{equation}
Using this similarity score, we can take advantage of MIPS to find the most relevant context to a query. 

In sum, Poly-DPR differs from Poly-Encoder in two major aspects:
(1) $K$ pre-computed representations of context as opposed to $K$ representations computed during inference, and (2) a faster similarity computation during inference.

\subsection{Hybrid Model}
\label{sec:hybrid}
In this paper, we also explore a hybrid model that combines the traditional approach of BM25 and neural retrievers.
We first retrieve the top-100 candidate articles using BM25 and a neural retriever (Poly-DPR) separately.
The scores produced by these two methods for each candidate are denoted by $S_\mathrm{BM25}$ and $S_\mathrm{NR}$ respectively and normalized to the $[0, 1]$ range to obtain $S_{BM25}^\prime$ and $S_{NR}^\prime$.
If a candidate article is not retrieved by a particular method, then its score for that method is $0$.
For each article, we get a new score:
\begin{equation}
    S_\mathrm{hybrid} = S_\mathrm{BM25}^\prime+ S_\mathrm{NR}^\prime.
    \label{eq:hybrid_score}
\end{equation}
Finally, we re-rank candidates based on $S_{hybrid}$ and pick the top candidates -- for BioASQ performance is evaluated on the top-10 retrieved candidates.

\section{Template Based Question Generation}\label{sec:qg}
\begin{figure}
    \centering
    \includegraphics[width=\linewidth]{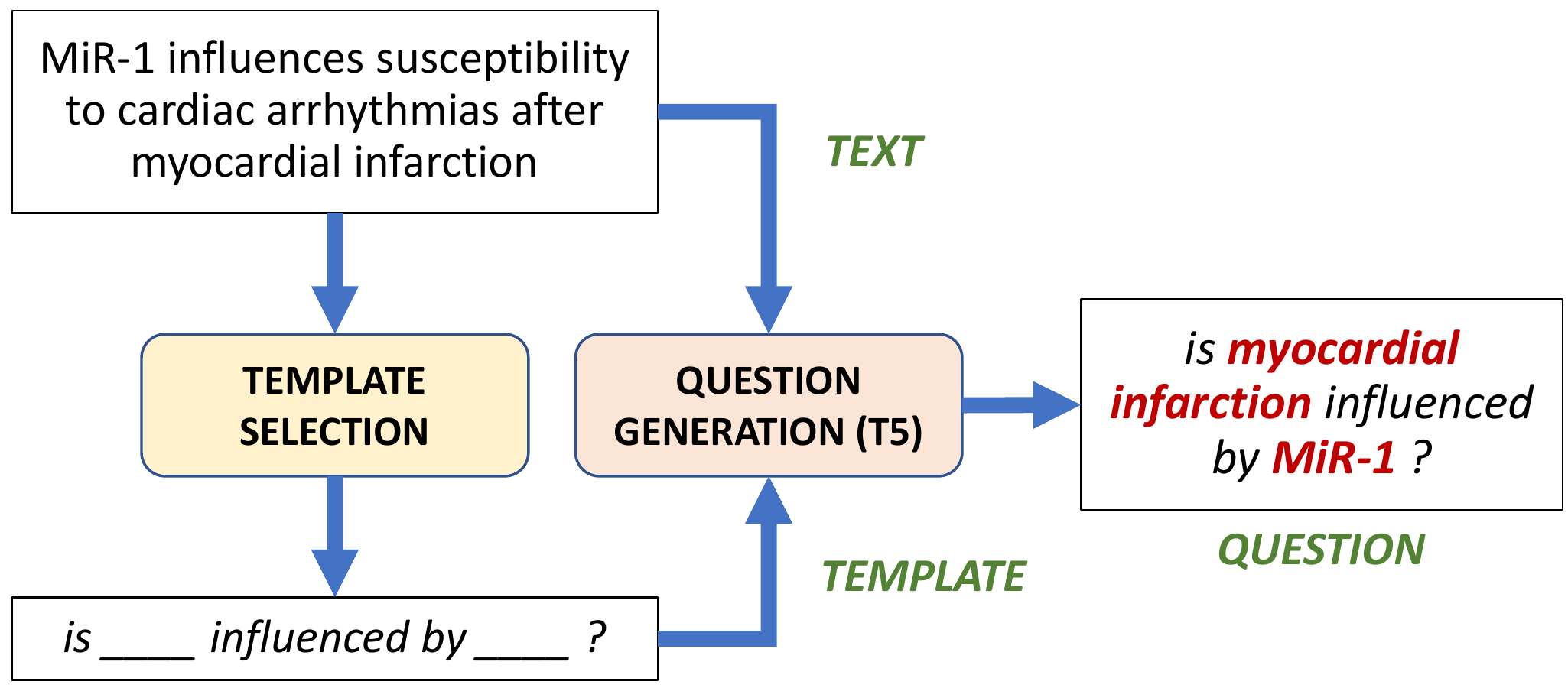}
    \caption{Overview of Template-Based Question Generation.}
    \label{fig:bionr_qg}
\end{figure}
\begin{figure*}[t]
    \centering
    \includegraphics[width=\linewidth]{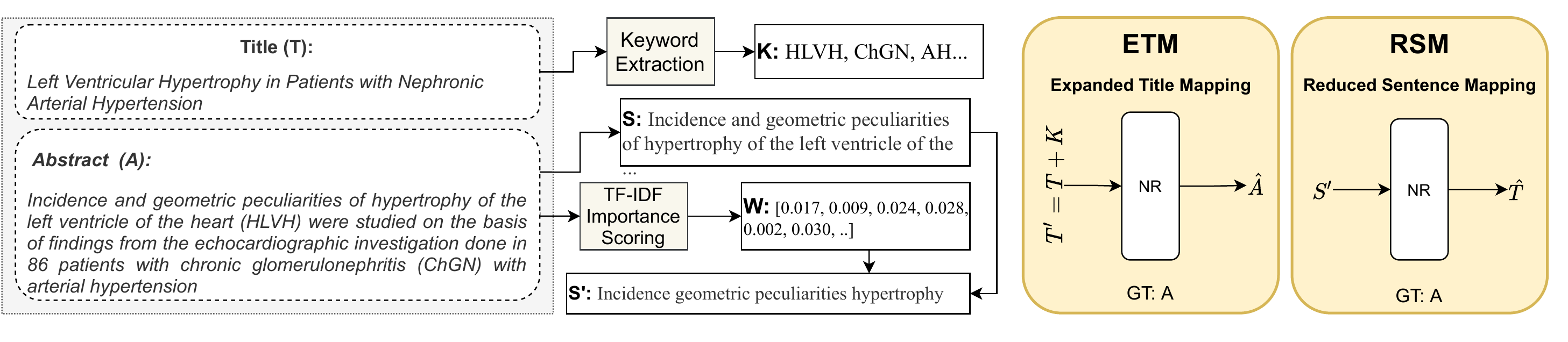}
    \caption{Poly-DPR is pre-trained on two novel tasks designed specifically for information retrieval applications. This figure illustrates the sample generation pipeline using the title and abstract from each sample in BioASQ.}
    \label{fig:pre-training}
\end{figure*}
We propose a template-based question generation approach -- \textit{TempQG}, that captures the style of the questions in the target domain. 
Our method consists of three modules: template extraction, template selection, and question generation. 

\paragraph{Template Extraction} aims to extract unique templates from which the questions in the training set can be generated. 
We first use bio-entity taggers from Spacy~\cite{spacy} to obtain a set of entities from the question.
We replace non-verb entities having a document frequency less than $k$ with an underscore (\_) -- this prevents common entities such as ``disease", ``gene" from being replaced. 
For e.g., given the question \textit{``Borden classification is used for which disease?''}, the entity tagger returns \textit{[``Borden classification'', ``disease'']}, but only the first entity clears our frequency-based criteria. 
As a result, the generated template is \textit{``\_ is used for which disease?''}.
This process gives us a preliminary list of templates. 
We then use a question similarity model (which returns a score between $[0,1]$) to compute the pairwise score between all templates.
Templates are assigned to a cluster if they have a minimum similarity of $0.75$
with existing templates of a cluster.
Once clusters have been formed, we choose either the sentence with the smallest or second-smallest length as the representative template.
These representative templates are used for question generation.

\paragraph{Template Selection.}
Given a text passage, we create a text-template dataset and train the PolyDPR architecture to retrieve a relevant template.
After the model is trained, we feed new text inputs to the model, obtain query encoding and compute the inner product with each template.
Templates with maximum inner product are selected to be used for QG.

\paragraph{Question Generation (QG).} 
We use a T5~\citep{Raffel2020ExploringTL} model for generating questions, by using text and template as conditional inputs.
To distinguish between these two inputs, we prepend each with the word ``template'' or ``context'', resulting in an input of the form:
$\{``template": template, ``context": text\}$. 
Figure~\ref{fig:bionr_qg} shows an illustrative example for the template-based question generation method abbreviated as \textit{TempQG}.
The context used for generating the questions are any two consecutive sentences in the abstract.
Given such a context, we first select 10 unique templates and concatenate each template with the context independently.
These are used by the question generation model to produce 10 initial questions;
duplicate questions are filtered out.

\section{Pre-training for Neural Retrieval}
\label{sec:pretrain-task}
Our aim is to design pre-training tasks specifically for the biomedical domain since documents in this domain bear the \textit{<title, abstract, main text>} structure of scientific literature.
This structure is not commonly found in documents such as news articles, novels, and text-books.
Domain-specific pre-training tasks have been designed by
\citet{Chang2020PretrainingTF} for Wikipedia documents which contains hyperlinks to other Wikipedia documents.
However most biomedical documents do not contain such hyperlinks, and a such, pre-training strategies recommended by \citet{Chang2020PretrainingTF} are incompatible with structure of biomedical documents.

Therefore, we propose Expanded Title Mapping (ETM) and Reduced Sentence Mapping (RSM), designed specifically for biomedical IR, to mimic the functionality required for open-domain question answering.
An overview is shown in Figure~\ref{fig:pre-training}.
The proposed tasks work for both short as well as long contexts.
In biomedical documents, each document has a title ($T$) and an abstract ($A$). 
We pre-train our models on ETM or RSM and then finetune them for retrieval.

\paragraph{Expanded Title Mapping (ETM).} 
For ETM, the model is trained to retrieve an abstract, given an extended title $T^\prime$ as a query.
$T^\prime$ is obtained by extracting top-$m$ keywords from the abstract based on the TF-IDF score, denoted as $K = \{k_1, k_2, \cdots, k_m\}$, and concatenating them with the title as: $T' = \{T, k_1, k_2, \cdots, k_m\}$. 
The intuition behind ETM is to train the model to match the main topic of a document (keywords and title) with the entire abstract.

\paragraph{Reduced Sentence Mapping (RSM).} 
RSM is designed to train the model to map a sentence from an abstract with the extended title $T^\prime$.
For a sentence $S$ from the abstract, we first get the weight of each word $W = \{ w_1, w_2, \cdots, w_n\}$ by the normalization of TF-IDF scores of each word.  
We then reduce $S$ to $S'$ by selecting the words with the top-$m$ corresponding weights.
The intuition behind a reduced sentence is to simulate a real query which usually is shorter than a sentence in a PubMed abstract. 
Furthermore, $S'$ includes important words based on the TF-IDF score, which is similar to a question including keywords.

\section{Experiments}\label{sec:exp}
\noindent{\bf Dataset.}
We focus on the document retrieval task in BioASQ8~\cite{Tsatsaronis2015AnOO} with a goal of retrieving a list of relevant documents to a question. 
This dataset contains $3234$ questions in the training set and five test sets \textit{(B1, B2, B3, B4, B5)} with $100$ questions each. Each question is equipped with a list of relevant documents and a list of relevant snippets of the documents.

\paragraph{Baselines.}
We compare our work with
BM25 and DPR in the short corpus setting, and BM25 and GenQ~\cite{Ma2021ZeroshotNP} in the large corpus setting. 
Note that the same test sets (B1-5) are used for evaluating both settings.
We also compare an alternative question generation method AnsQG~\citep{Chan2019ARB} in which an answer extraction model first extracts an answer from a context and a question generation model uses the answer as well as the text to generate a question.
Similarly, we compare our method with an existing pre-training task ICT~\citep{Lee2019LatentRF}. 
Retrieval systems for the BioASQ task typically follow a two-step process: retrieval of candidates and re-ranking.
The focus of this paper is on improving the former, and thus we use different retrieval methods as baselines and do not compare with state-of-the-art systems that use various re-ranking methods.
We use Mean Average Precision (MAP) as our evaluation metric.

\subsection{Experimental Settings}
\paragraph{Size of Corpus.}
PubMed is a large corpus containing 19 million articles,
each with a title and an abstract. 
Due to this large corpus size, indexing the entire corpus takes a significantly long time. 
To conduct comprehensive experiments and to efficiently evaluate the impacts of each proposed method, we construct a small corpus with $133,084$ articles in total: $33,084$ articles belonging to the training and test sets of BioASQ8, and an additional $100K$ articles that are randomly sampled from the entire corpus. 
We compare and analyze the impacts of different techniques on the small corpus setting, and then evaluate the best model on the large corpus setting. 

\paragraph{Length of Context.}
We use two context lengths for training neural retrievers as well as for indexing the corpus: $128$ (short) and $256$ (long).
For indexing the corpus, each article is segmented into multiple pieces using a window length of two sentences and stride of one sentence. 
We use document snippets as contexts in the short context setting and the entire document in the long context setting.
With our generated training dataset (TempQG), given a question, in the short context setting, we use sentences which were used for generating the question as the training data, whereas in the long context setting we use the source document from which the question was generated as the context. 
We use RSM as the pre-training task for short contexts and either ETM or ICT with long contexts.

\paragraph{Training Setup.}
For all experiments, we use BioBERT~\citep{Lee2020BioBERTAP} as the initial model for both query and context encoders.
For BM25, we use an implementation from Pyserini~\citep{Lin2021PyseriniAE} with default hyperparameters $k{=}0.9$ and $b{=}0.4$. 
We also try $k{=}1.2$ and $b{=}0.75$ as used by~\citet{Ma2021ZeroshotNP} and find that the default setting is slightly better.
For Poly-DPR, the number of representations $K$ is set as $6$ after a hyper-parameter search.
While larger values of $K$ improve results, it makes indexing slower. 
Details about the hyper-parameter search are provided in the appendix.

\subsection{Results}
\begin{table}[t]
    \centering
    \resizebox{\linewidth}{!}{
    \begin{tabular}{@{}cllcccccc@{}}
        \toprule
        \textbf{CL} & \textbf{PT} & \textbf{FT} & \textbf{B1} & \textbf{B2} & \textbf{B3} & \textbf{B4} & \textbf{B5} & \textbf{Avg.}\\
        \toprule
    \multirow{5}{*}{\rotatebox[origin=c]{90}{Short (128) }} 
    & -        & BioASQ       &54.48      &50.51       & 53.8       &59.06     &48.71     & 53.31 \\
    & -   & TempQG  &62.92  &58.79 &62.94 & 70.30 &63.39 & 63.67  \\
    & RSM      & BioASQ       &65.94      &57.43       & 61.89     &69.01    &58.23      & 62.50  \\
    & RSM & AnsQG & 56.84 & 55.79 & 57.52 & 58.68 & 55.15 & 56.80\\
    & RSM & TempQG  & 64.71 & 64.92 & 64.28  & 73.11  &  66.29 & 66.66 \\
    \midrule
    \multirow{8}{*}{\rotatebox[origin=c]{90}{Long (256)}}
    & - & BioASQ &35.69 &32.66  &32.26 &38.28  &30.87 & 33.95  \\
    & - & TempQG &63.95  & 59.51 &62.98 &66.71  &62.8 &63.19  \\
    & ICT~\shortcite{Lee2019LatentRF} & BioASQ &54.44  &47.37  &52.61 &53.69  &44.38 &50.50  \\
    & ETM & BioASQ & 56.63 &46.63 &52.79 &56.97  &49.61 &52.53  \\
    & ETM & TempQG & 64.57 &58.51 & 64.02  &68.44 &62.6 & 63.62  \\
    & ETM & AnsQG  & 54.44 & 49.95 & 48.42 & 58.15 & 52.6 & 52.71\\
    & ICT+ETM & BioASQ &51.33 &49.43 &49.36 &53.19 &43.58 &49.38  \\
    & ICT+ETM & TempQG & {64.93} &58.49  &60.18  & 69.42 & 64.87 &63.58  \\
    \bottomrule
    \end{tabular}
    }
    \caption{Effect of pre-training tasks (PT) and fine-tuning datasets (FT) on the performance of Poly-DPR with two context lengths (CL) on the BioASQ small corpus test set. 
    }
    \label{tab:pt_ablation} 
\end{table}
        
\label{sec:main-result}
{\bf Effect of Pre-Training Tasks and Fine-Tuning Datasets.}
Table~\ref{tab:pt_ablation} shows results when Poly-DPR is trained with different methods of pre-training and different fine-tuning datasets.
We see that both RSM and ETM lead to improvements even when the finetuning task has only a limited amount of supervised data, i.e. BioASQ. 
When compared to Poly-DPR trained without any pre-training, RSM improves by ${\sim}9\%$ and ETM by ${\sim}18\%$.
ETM is better than the existing pre-training method ICT~\citep{Lee2019LatentRF} by ${\sim}2\%$. 
When the size of fine-tuning set is large, i.e.\ with our question generation method (TempQG), the gains due to pretraining are higher with short contexts than with large contexts. 
We believe this to be a result of the finetuning dataset in the long-context setting being significantly larger than the pre-training dataset, thereby having a larger effect on the training process.
Dataset statistics for pretraining are in the Appendix.

We also see that when Poly-DPR is only trained on BioASQ, the performance with small contexts is much better than with long contexts ($53.31\%$ vs $33.95\%$). This suggests that Poly-DPR trained on the small corpus finds it difficult to produce robust representations for long contexts. 
On the other hand, the performance of Poly-DPR variants trained on TempQG is close for short and long contexts, which suggests that large-scale relevant training data improves representations.

\begin{table}[t]
    \centering
    \resizebox{\linewidth}{!}{
    \begin{tabular}{@{}lcccccc@{}}
        \toprule
        \textbf{ Model} & {\bf B1}   & {\bf B2} & {\bf B3} & {\bf B4} & {\bf B5} & {\bf Avg.}\\
        \midrule
        \multicolumn{7}{c}{\it Small Corpus}\\
        BM25~\shortcite{Robertson2009ThePR} &62.15 & 61.30 &66.62 &74.14 &61.30 & 65.10 \\
        DPR$_{128}$~\shortcite{Karpukhin2020DensePR} & 54.48 & 50.51 & 53.80 & 59.06 & 48.71 & 53.31\\
        DPR$_{256}$               & 44.86 & 41.18 & 40.25 & 47.78 & 40.42 & 42.89\\
        Hybrid (BM25+DPR$_{128}$)  & \textbf{66.55} & 61.29 & 68.08  &72.91  &60.30 & 65.83 \\
        \textit{Poly-DPR$_{128}$ (Ours)}   & 64.71 & \textbf{64.92} & 64.28 & 73.11 & \textbf{66.29} & 66.66 \\
        \textit{Poly-DPR$_{256}$ (Ours)}   & 64.57 & 58.51 & 64.02 & 68.44 & 62.60 & 63.62 \\
        \textit{Hybrid (BM25+Poly-DPR$_{128}$)} &66.30  &64.90  &\textbf{69.54}  &\textbf{75.71} &64.82 &\textbf{68.25} \\
        \midrule
        \multicolumn{7}{c}{\it Large Corpus}\\
        BM25 &28.50 &27.82  &37.97  &41.91  &35.42  &34.32 \\
        GenQ~\shortcite{Ma2021ZeroshotNP} & 28.90 & 20.30 & 30.70 & 29.00 & 33.10 & 28.40\\
        \textit{Poly-DPR$_{128}$ (Ours)} & \textbf{35.10} & 29.07 & 32.74 & 33.31 & 35.54 & 33.15\\
        \textit{Hybrid (BM25+Poly-DPR$_{128}$)} & 30.02 & \textbf{31.31} & \textbf{39.79} & \textbf{42.18}  & \textbf{37.99}  & \textbf{36.26} \\
        \midrule
        \multicolumn{7}{c}{\it Large Corpus SOTA (Re-ranking)}\\
        PA\shortcite{kazaryan2020transformer} &35.91 & \textbf{39.45} &52.73 &41.15 &\textbf{52.02} & 44.25 \\
        bioinfo-4~\shortcite{almeida2020bit} &\textbf{38.23} & 36.86 &51.08 &46.77 &50.98 & \textbf{44.78} \\
        AUEB-4~\shortcite{pappas2020aueb} &5.47 & 7.23 &\textbf{53.29} &\textbf{49.92} &49.53 & 33.09 \\
        \bottomrule
    \end{tabular}
    }
    \caption{
    Comparison between our method (Poly-DPR) with baseline methods in the small corpus and large corpus settings.
    The bottom section shows performance of existing methods that make improvements in the re-ranking method.
    }
    \label{tab:results_bioasq}
\end{table}
\begin{table*}[t]
    \centering 
    \small
    \resizebox{\linewidth}{!}{
    \begin{tabular}{@{}lp{0.68\linewidth}p{0.1\linewidth}p{0.25\linewidth}@{}}
        \toprule 
        \textbf{\#} &\textbf{Context} & \textbf{Template} & \textbf{Generated Question} \\
        \toprule
        1 & The lysosomal-membrane protein type 2A (LAMP-2A) acts as the receptor for the substrates of chaperone-mediated autophagy (CMA), which should undergo unfolding before crossing the lysosomal membrane and reaching the lumen for degradation.
            & which receptor is targeted by \_ 
            & Which receptor is targeted by LAMP-2A? \\
        \midrule
        2 & Is Tokuhashi score suitable for evaluation of life expectancy before surgery in Iranian patients with spinal metastases? One of the most important selection criteria for spinal metastases surgery is life expectancy and the most important system for this prediction has been proposed by Tokuhashi.
            & what is evaluated with \_ 
            & What is the Tokuhashi score? \\
        \midrule
        
        3 & Lambert-Eaton myasthenic syndrome (LEMS) is a pre-synaptic disorder of the neuromuscular and autonomic transmission mediated by antibodies to voltage-gated calcium channels at the motor nerve terminal. 
            & \_ is diagnosed in which \_ 
            & Lambert-Eaton myasthenic syndrome is diagnosed in which neuromuscular and autonomic pathways? \\
        \bottomrule
    \end{tabular}
    }
    \caption{Ilustrative examples for templates and questions generated by TempQG. More examples are in the Appendix.
    }
    \label{tab:ques-example}
\end{table*}
\begin{table}[t]
    \centering
    \resizebox{\linewidth}{!}{
    \begin{tabular}{@{}lrccccccc@{}}
    \toprule
    {\bf Index Unit} & {\bf Mem.} & {\bf Time} &  {\bf B1}  & {\bf B2}  & {\bf B3}  & {\bf B4}  & {\bf B5} & {\bf Avg.}\\

    \toprule
    2-sents & 21.0 G & 321   &64.71& {\bf 64.92} &{\bf 64.28}  &{\bf 73.11}  & {\bf 66.29} &{\bf 66.66} \\
    128-chunk & 8.1 G & 206  &{\bf 65.16} & 63.24 &63.72  &72.13  & 65.29 & 65.91 \\
    256-chunk & 4.5 G & 192  & 63.76 & 59.71 &62.70 &67.21  & 64.17 & 63.51 \\
    Full & 2.8 G & 101  & 61.92 & 57.84 &60.01 &61.11  & 62.66 & 60.71 \\
    \midrule
    2-sents & 21.0 G & 321  & {\bf 64.65} & \bf59.21 & 63.65  &\bf70.90 &\bf65.97 & \bf64.88  \\
    128-chunk & 8.1 G & 206  & 64.11 & 58.08 & \bf64.15  &69.90 & 63.16 & 63.88 \\
    256-chunk & 4.5 G & 192  &64.57 &58.51 & 64.02  &68.44 &62.6 & 63.62  \\
    Full & 2.8 G & 101 &60.06 &56.38 & 61.99  &65.01 &59.63 & 60.61 \\
    
    \toprule
    \end{tabular}
    }
    \caption{Two best NR models in short and long context: the first block is Poly-DPR pretrained with RSM and fine-tuned on TempQG (short); the second block is Poly-DPR pretrained with ETM and fine-tuned on TempQG (long). }
    \label{tab:indexing}
\end{table}
\begin{table}[t]
    \centering 
    \small
    \resizebox{\linewidth}{!}{
    \begin{tabular}{@{}ccccccc@{}}
        \toprule
         {\bf \# Templates}  & {\bf B1}   & {\bf B2} & {\bf B3} & {\bf B4} & {\bf B5} & {\bf Average}\\
         \toprule
            1  & {\bf 67.21} & 62.43 & \textbf{66.49} & {\bf 72.15} & 61.55 & 65.96 \\
            5  & 66.76 & 62.19 & 66.41 & 71.55 & {\bf 64.33} & 66.25 \\
            10 & 64.71   & {\bf 64.92} & 64.28 & 73.11 & 62.29 & {\bf 66.66} \\
        \toprule
        \end{tabular}
        }
    \caption{Effect of number of templates on performance.
    }
\label{tab:num_templates}
\end{table}
\paragraph{Comparison with Baselines.}
Table~\ref{tab:results_bioasq} shows a comparison between baselines and our best model (Poly-DPR with short context (128) pre-trained with RSM and finetuned on TempQG).
Note that our model is only trained on datasets acquired from the small corpus.
However, we evaluate the same model on the large corpus test set.

In the \textbf{small corpus setting}, it can be seen that our model outperforms all existing methods in the small corpus setting, and is better than DPR by $13.3\%$ and $20.8\%$ in short (128) and long (256) context lengths respectfully. 
In the \textbf{large corpus setting}, our method is better than GenQ~\citep{Ma2021ZeroshotNP} on all five test sets.
This shows that our method, which uses 10 million generated samples is better than GenQ which uses 83 million samples for training, thus showing the effectiveness of our template-based question generation method.
Although our method performs better than BM25 on B1, B2, B5, the average performance is slightly worse ($-1.17\%$). 
For the hybrid method, we apply our best Poly-DPR model to index the entire corpus, and use the procedure as described in Sec~\ref{sec:hybrid}.
Our hybrid method which combines BM25 and Poly-DPR, is better than all existing methods.

We also report state-of-the-art (SOTA) results reported on the BioASQ8 leader-board \footnote{\url{http://participants-area.bioasq.org/results/8b/phaseA/}}, for methods -- PA~\citep{kazaryan2020transformer}, bioinf-4~\citep{almeida2020bit}, and AUEB-4~\citep{pappas2020aueb}.
These approaches are a combination of retrieval and improved re-ranking methods.
Since this paper is concerned with improving retrieval and does not study re-ranking, we do not compare our methods directly with these approaches, but report them for completeness.


\begin{table*}[t]
    \centering 
    \small
    \resizebox{\linewidth}{!}{
    \begin{tabular}{@{}lp{0.3\linewidth}p{0.72\linewidth}@{}}
        \toprule 
        & \textbf{Question} & \textbf{Explanation} \\
        \toprule
        B1 &  What is minodixil approved for?& minodixil is a typo, the correct one is minoxidil\\
        B2 & List 5 proteins with antioxidant properties?  & BM25 fails to connect proteins and antioxidant properties, and retrieves documents all related to antioxidant, however, they are not about proteins nor antioxidant proteins.\\
        B3 & How  large  is  a  lncRNAs? & BM25 retrieves document about lncRNAs but not about how large it is. \\
        \midrule
        P1 & What is Xanamem? & NR fails to retrieve any document related to Xanamem, rather, it retrieves documents that lexical similar to Xanamem such as Ximenia, Xadago, and Xenopus.\\
        P2 & Does an interferon (IFN) signature exist for SLE patients? & NR ranks documents about interferon higher than documents of SLE patients and documents of both. In the retrieved documents, interferon appears rather frequently.\\
        \bottomrule
    \end{tabular}
    }
    \caption{
    Examples of the common failure modes of BM25 and Poly-DPR.
    More examples are in the Appendix.
    }
    \label{tab:error_analysis}
\end{table*}

\subsection{Analysis}
We provide ablation studies and analysis of the effect of our design choices and hyper-parameters on model performance.
All results are reported on the small corpus. 

\paragraph{Granularity of Indexing.}
Here we examine the impact of indexing units. 
We conjecture that the representation produced with a shorter indexing unit is better than the one with a longer indexing unit, and thus an NR should perform better if the indexing unit is short. 
To verify this, we use our best Poly-DPR models that are trained in short and long context settings. 
We compare four indexing units, 
\begin{itemize}[nosep,noitemsep]
    \item \textit{2-sents}:  two consecutive sentences, 
    \item \textit{128 chunk}: a chunk with maximum length of 128 tokens that includes multiple consecutive sentences,
    \item \textit{256 chunk}: a chunk with maximum length of 256 tokens that includes multiple consecutive sentences, and 
    \item \textit{Full}: the entire article including title and abstract, and we use 512 tokens to encode each article. 
\end{itemize}
The results are shown in Table \ref{tab:indexing}; we see that the smaller indexing units yield better performance, even for the model that is trained in long context setting.
We also present the memory (\textit{Mem.}) and inference time (\textit{Time}) which depend upon the choice of indexing unit. 
The inference time refers to the number of seconds taken to retrieve 10 documents for 100 questions.
Table \ref{tab:indexing} shows that a smaller indexing unit requires more memory and longer inference time. 
Thus, there is a trade-off between retrieval quality and memory as well as inference time.
Future work could explore ways to improve the efficiency of neural retrievers to mitigate this trade-off.

\paragraph{Number of Templates for Generating Questions}
We study three values for the number of templates, 1, 5, and 10, and report the results for Poly-DPR in Table~\ref{tab:kvalue}.
We see that training Poly-DPR on questions generated from one template is already better than BM25. 
While increasing the number of templates yields better performance, the improvement is relatively small, and we conjecture that this could be due to lower-quality or redundant templates.
A question filtering module can be used to control the quality of the questions as shown in previous work~\citep{Alberti2019SyntheticQC}.

\paragraph{Number of Context Representations} 
\begin{table}[t]
    \small
    \centering
    \resizebox{0.82\linewidth}{!}{
    \begin{tabular}{@{}ccccccc@{}}
        \toprule
         {\bf K} & {\bf B1}   & {\bf B2} & {\bf B3} & {\bf B4} & {\bf B5} & {\bf Average}\\
         \toprule
             0  & 62.06  & {\bf 61.81}& 61.85 & 66.69 & 61.30 & 62.74\\
            6  & 62.92 & 58.79 & {\bf 62.94} & 70.30 & 63.39 & 63.67\\
            12  & {\bf 65.22} & 60.86 & 62.59 & {\bf 70.50} & {\bf 66.21} & {\bf 65.08}\\
            \midrule
            0  & 61.70  & 58.28 & 58.62 & 67.33 & 61.48 & 61.48\\
            6  & {\bf 63.95} & {\bf 59.51} & {\bf 62.98} & 66.71 & 62.80 & 63.19\\
            12 & 63.83 & 57.81 & 62.72 & {\bf 70.00} & {\bf 63.64} & {\bf 63.60}\\
        \toprule
    \end{tabular}
    }
\caption{Comparison among different values of K for Poly-DPR in both short and long context settings.}
\label{tab:kvalue}
\end{table}
Poly-DPR encodes a context into $K$ vector representations.
We study the effect of three values of $K$ (0, 6, and 12) on model performance, both with short (128) and long (256) contexts.
All models are trained directly on the TempQG without pretraining.
Table~\ref{tab:kvalue} shows that a larger K value yields better performance. 
This observation is aligned with \citet{Humeau2020PolyencodersAA}.

\subsection{Question Generation Analysis}
Table \ref{tab:ques-example} shows examples of selected templates and generated questions. 
We see that our template-based generation approach can produce diverse and domain-style questions. 
In particular, the template selection model can select relevant templates by word matching and semantic matching. Given a template and a context, the question generator employs three ways to generate questions. 
\begin{enumerate}[nosep,noitemsep]
    \item {\bf Fill in the blank}: the generator fills the blank in the template by key entities mentioned in the context without changing the template, as shown by Example 1. 
    \item {\bf Changing partially}: the generator produces questions by using part of the template and ignores some irrelevant part  as shown by Example 2. 
    \item {\bf Ignoring entirely}: the generator ignores the template entirely and generates questions that are not relevant to the given context as shown by Example 3.
\end{enumerate}

\subsection{Error Analysis}
To better understand the differences between BM25 and NR, we study their failure modes.
From the BioASQ test set, we select questions on which either BM25 or Poly-DPR perform poorly, and categorize these failure cases.

\paragraph{Failures Cases of BM25.} 
We found 91 failure cases on which the MAP score of BM25 is 0 for 41 cases, and the performance of BM25 is at least 0.5 less than Poly-DPR for 50 cases. 
Upon manual inspection, we identify three common categories of these failures:
\begin{enumerate}[label=(\textbf{B\arabic*}),nosep,noitemsep,leftmargin=*]
    \item Questions contain keywords with typographical errors,
    \item Questions mention multiple entities related to each other. BM25 may fail to retrieve documents that connect these entities.
    \item Questions mention conceptual properties of entities and answers are values. For example, "how large" is a conceptual property and "200" is the answer value. Although BM25 retrieves documents related to the entities in questions, those may not contain the answer.
\end{enumerate}

\paragraph{Failure cases of Poly-DPR.} 
There we $55$ failure cases of Poly-DPR, including 23 cases with 0 MAP score and 32 case where the score for BM25 is at least 0.5 better than Poly-DPR.  
The common failure modes of Poly-DPR are:
\begin{enumerate}[label=(\textbf{P\arabic*}),nosep,noitemsep,leftmargin=*]
    \item Questions are simple but focused on rare entities which Poly-DPR fails to retrieve.
    This conforms with the finding that NR performs significantly worse than BM25 on entity-questions~\citep{sciavolino2021simple}. 
    We find that for such questions, retrieved entities and entities in the question are lexical similar or have overlapping substrings, which in turn could be due to the WordPiece embeddings~\citep{wu2016google} used in BERT. 
    \item Questions mention multiple entities. Articles that contain frequent entities are ranked higher than articles that include \textit{all} entities in the question.
\end{enumerate}

\section{Discussion and Conclusion}\label{sec:conclude}
In this work, we show that DPR, a neural retriever, which achieves better results than BM25 on retrieval-based tasks in several open-book question answering benchmarks, is unable to surpass BM25 on biomedical benchmarks such as BioASQ. 
We address this drawback of NRs with a three-pronged approach with Poly-DPR: a new model architecture, TempQG: a template-based question generation method, and two new pre-training tasks designed for biomedical documents.
TempQG can generate high quality domain-relevant questions which positively impact downstream performance. 
While in this paper, we apply TempQG to a small corpus of 100,000 PubMed articles, we show that this method can surpass neural retrievers when trained on small or large corpora.
Our model achieves better performance than BM25 in the small corpus setting, but it falls short by ${\sim}1\%$ in the large corpus setting.
However, we show that a hybrid model combining our approach and BM25 is better than all previous baselines on the entire corpus.
In the future, applying our question generation methods to the entire PubMed corpus, and combining our approach with improved re-ranking techniques could potentially result in further improvement.

\section*{Acknowledgements}
This research is based upon work supported in part by the National Science Foundation under Grant No. 1816039 and the Office of Naval Research under grant N00014-20-1-2332.
Any opinions, findings, or recommendations expressed in this material are those of the authors and do not necessarily reflect the views of the supporting agencies.
\bibliography{reference}

\begin{thebibliography}{42}
\providecommand{\natexlab}[1]{#1}

\bibitem[{Alberti et~al.(2019)Alberti, Andor, Pitler, Devlin, and
  Collins}]{Alberti2019SyntheticQC}
Alberti, C.; Andor, D.; Pitler, E.; Devlin, J.; and Collins, M. 2019.
\newblock Synthetic {QA} Corpora Generation with Roundtrip Consistency.
\newblock In \emph{Proceedings of the 57th Annual Meeting of the Association
  for Computational Linguistics}, 6168--6173. Florence, Italy: Association for
  Computational Linguistics.

\bibitem[{Almeida and Matos(2020)}]{almeida2020bit}
Almeida, T.; and Matos, S. 2020.
\newblock BIT. UA at BioASQ 8: Lightweight Neural Document Ranking with
  Zero-shot Snippet Retrieval.
\newblock In \emph{CLEF (Working Notes)}.

\bibitem[{Banerjee, Gokhale, and Baral(2021)}]{banerjee2021self}
Banerjee, P.; Gokhale, T.; and Baral, C. 2021.
\newblock Self-Supervised Test-Time Learning for Reading Comprehension.
\newblock In \emph{Proceedings of the 2021 Conference of the North American
  Chapter of the Association for Computational Linguistics: Human Language
  Technologies}, 1200--1211.

\bibitem[{Banerjee et~al.(2021)Banerjee, Gokhale, Yang, and
  Baral}]{banerjee2021weaqa}
Banerjee, P.; Gokhale, T.; Yang, Y.; and Baral, C. 2021.
\newblock WeaQA: Weak supervision via captions for visual question answering.
\newblock In \emph{Findings of the Association for Computational Linguistics:
  ACL-IJCNLP 2021}, 3420--3435.

\bibitem[{Chan and Fan(2019)}]{Chan2019ARB}
Chan, Y.-H.; and Fan, Y.-C. 2019.
\newblock A Recurrent {BERT}-based Model for Question Generation.
\newblock In \emph{Proceedings of the 2nd Workshop on Machine Reading for
  Question Answering}, 154--162. Hong Kong, China: Association for
  Computational Linguistics.

\bibitem[{Chang et~al.(2020)Chang, Yu, Chang, Yang, and
  Kumar}]{Chang2020PretrainingTF}
Chang, W.; Yu, F.~X.; Chang, Y.; Yang, Y.; and Kumar, S. 2020.
\newblock Pre-training Tasks for Embedding-based Large-scale Retrieval.
\newblock In \emph{8th International Conference on Learning Representations,
  {ICLR} 2020, Addis Ababa, Ethiopia, April 26-30, 2020}. OpenReview.net.

\bibitem[{Devlin et~al.(2019)Devlin, Chang, Lee, and
  Toutanova}]{Devlin2019BERTPO}
Devlin, J.; Chang, M.-W.; Lee, K.; and Toutanova, K. 2019.
\newblock {BERT}: Pre-training of Deep Bidirectional Transformers for Language
  Understanding.
\newblock In \emph{Proceedings of the 2019 Conference of the North {A}merican
  Chapter of the Association for Computational Linguistics: Human Language
  Technologies, Volume 1 (Long and Short Papers)}, 4171--4186. Minneapolis,
  Minnesota: Association for Computational Linguistics.

\bibitem[{Dong et~al.(2019)Dong, Yang, Wang, Wei, Liu, Wang, Gao, Zhou, and
  Hon}]{Dong2019UnifiedLM}
Dong, L.; Yang, N.; Wang, W.; Wei, F.; Liu, X.; Wang, Y.; Gao, J.; Zhou, M.;
  and Hon, H. 2019.
\newblock Unified Language Model Pre-training for Natural Language
  Understanding and Generation.
\newblock In Wallach, H.~M.; Larochelle, H.; Beygelzimer, A.;
  d'Alch{\'{e}}{-}Buc, F.; Fox, E.~B.; and Garnett, R., eds., \emph{Advances in
  Neural Information Processing Systems 32: Annual Conference on Neural
  Information Processing Systems 2019, NeurIPS 2019, December 8-14, 2019,
  Vancouver, BC, Canada}, 13042--13054.

\bibitem[{Du and Cardie(2018)}]{Du2018HarvestingPQ}
Du, X.; and Cardie, C. 2018.
\newblock Harvesting Paragraph-level Question-Answer Pairs from {W}ikipedia.
\newblock In \emph{Proceedings of the 56th Annual Meeting of the Association
  for Computational Linguistics (Volume 1: Long Papers)}, 1907--1917.
  Melbourne, Australia: Association for Computational Linguistics.

\bibitem[{Fabbri et~al.(2020)Fabbri, Ng, Wang, Nallapati, and
  Xiang}]{Fabbri2020TemplateBasedQG}
Fabbri, A.; Ng, P.; Wang, Z.; Nallapati, R.; and Xiang, B. 2020.
\newblock Template-Based Question Generation from Retrieved Sentences for
  Improved Unsupervised Question Answering.
\newblock In \emph{Proceedings of the 58th Annual Meeting of the Association
  for Computational Linguistics}, 4508--4513. Online: Association for
  Computational Linguistics.

\bibitem[{Guu et~al.(2020)Guu, Lee, Tung, Pasupat, and Chang}]{Guu2020REALMRL}
Guu, K.; Lee, K.; Tung, Z.; Pasupat, P.; and Chang, M.-W. 2020.
\newblock REALM: Retrieval-Augmented Language Model Pre-Training.
\newblock \emph{ArXiv}, abs/2002.08909.

\bibitem[{Honnibal et~al.(2017)Honnibal, Montani, Van~Landeghem, and
  Boyd}]{spacy}
Honnibal, M.; Montani, I.; Van~Landeghem, S.; and Boyd, A. 2017.
\newblock SpaCy: Industrial-strength Natural Language Processing in Python.

\bibitem[{Hosking and Lapata(2021)}]{hosking-lapata-2021-factorising}
Hosking, T.; and Lapata, M. 2021.
\newblock Factorising Meaning and Form for Intent-Preserving Paraphrasing.
\newblock In \emph{Proceedings of the 59th Annual Meeting of the Association
  for Computational Linguistics and the 11th International Joint Conference on
  Natural Language Processing (Volume 1: Long Papers)}, 1405--1418. Online:
  Association for Computational Linguistics.

\bibitem[{Humeau et~al.(2020)Humeau, Shuster, Lachaux, and
  Weston}]{Humeau2020PolyencodersAA}
Humeau, S.; Shuster, K.; Lachaux, M.; and Weston, J. 2020.
\newblock Poly-encoders: Architectures and Pre-training Strategies for Fast and
  Accurate Multi-sentence Scoring.
\newblock In \emph{8th International Conference on Learning Representations,
  {ICLR} 2020, Addis Ababa, Ethiopia, April 26-30, 2020}. OpenReview.net.

\bibitem[{Joshi et~al.(2017)Joshi, Choi, Weld, and
  Zettlemoyer}]{Joshi2017TriviaQAAL}
Joshi, M.; Choi, E.; Weld, D.; and Zettlemoyer, L. 2017.
\newblock {T}rivia{QA}: A Large Scale Distantly Supervised Challenge Dataset
  for Reading Comprehension.
\newblock In \emph{Proceedings of the 55th Annual Meeting of the Association
  for Computational Linguistics (Volume 1: Long Papers)}, 1601--1611.
  Vancouver, Canada: Association for Computational Linguistics.

\bibitem[{Karpukhin et~al.(2020)Karpukhin, Oguz, Min, Lewis, Wu, Edunov, Chen,
  and Yih}]{Karpukhin2020DensePR}
Karpukhin, V.; Oguz, B.; Min, S.; Lewis, P.; Wu, L.; Edunov, S.; Chen, D.; and
  Yih, W.-t. 2020.
\newblock Dense Passage Retrieval for Open-Domain Question Answering.
\newblock In \emph{Proceedings of the 2020 Conference on Empirical Methods in
  Natural Language Processing (EMNLP)}, 6769--6781. Online: Association for
  Computational Linguistics.

\bibitem[{Kazaryan, Sazanovich, and Belyaev(2020)}]{kazaryan2020transformer}
Kazaryan, A.; Sazanovich, U.; and Belyaev, V. 2020.
\newblock Transformer-Based Open Domain Biomedical Question Answering at
  BioASQ8 Challenge.
\newblock In \emph{CLEF (Working Notes)}.

\bibitem[{Khattab and Zaharia(2020)}]{Khattab2020ColBERTEA}
Khattab, O.; and Zaharia, M. 2020.
\newblock ColBERT: Efficient and Effective Passage Search via Contextualized
  Late Interaction over {BERT}.
\newblock In Huang, J.; Chang, Y.; Cheng, X.; Kamps, J.; Murdock, V.; Wen, J.;
  and Liu, Y., eds., \emph{Proceedings of the 43rd International {ACM} {SIGIR}
  conference on research and development in Information Retrieval, {SIGIR}
  2020, Virtual Event, China, July 25-30, 2020}, 39--48. {ACM}.

\bibitem[{Krishna, Bernstein, and Fei{-}Fei(2019)}]{krishna2019information}
Krishna, R.; Bernstein, M.; and Fei{-}Fei, L. 2019.
\newblock Information Maximizing Visual Question Generation.
\newblock In \emph{{IEEE} Conference on Computer Vision and Pattern
  Recognition, {CVPR} 2019, Long Beach, CA, USA, June 16-20, 2019}, 2008--2018.
  Computer Vision Foundation / {IEEE}.

\bibitem[{Kwiatkowski et~al.(2019)Kwiatkowski, Palomaki, Redfield, Collins,
  Parikh, Alberti, Epstein, Polosukhin, Devlin, Lee, Toutanova, Jones, Kelcey,
  Chang, Dai, Uszkoreit, Le, and Petrov}]{Kwiatkowski2019NaturalQA}
Kwiatkowski, T.; Palomaki, J.; Redfield, O.; Collins, M.; Parikh, A.; Alberti,
  C.; Epstein, D.; Polosukhin, I.; Devlin, J.; Lee, K.; Toutanova, K.; Jones,
  L.; Kelcey, M.; Chang, M.-W.; Dai, A.~M.; Uszkoreit, J.; Le, Q.; and Petrov,
  S. 2019.
\newblock Natural Questions: A Benchmark for Question Answering Research.
\newblock \emph{Transactions of the Association for Computational Linguistics},
  7: 452--466.

\bibitem[{Lee et~al.(2020)Lee, Yoon, Kim, Kim, Kim, So, and
  Kang}]{Lee2020BioBERTAP}
Lee, J.; Yoon, W.; Kim, S.; Kim, D.; Kim, S.; So, C.~H.; and Kang, J. 2020.
\newblock BioBERT: a pre-trained biomedical language representation model for
  biomedical text mining.
\newblock \emph{Bioinformatics}, 36: 1234 -- 1240.

\bibitem[{Lee, Chang, and Toutanova(2019)}]{Lee2019LatentRF}
Lee, K.; Chang, M.-W.; and Toutanova, K. 2019.
\newblock Latent Retrieval for Weakly Supervised Open Domain Question
  Answering.
\newblock In \emph{Proceedings of the 57th Annual Meeting of the Association
  for Computational Linguistics}, 6086--6096. Florence, Italy: Association for
  Computational Linguistics.

\bibitem[{Lewis, Denoyer, and Riedel(2019)}]{lewis2019unsupervised}
Lewis, P.; Denoyer, L.; and Riedel, S. 2019.
\newblock Unsupervised Question Answering by Cloze Translation.
\newblock In \emph{Proceedings of the 57th Annual Meeting of the Association
  for Computational Linguistics}, 4896--4910. Florence, Italy: Association for
  Computational Linguistics.

\bibitem[{Lewis, Stenetorp, and Riedel(2021)}]{Lewis2021QuestionAA}
Lewis, P.; Stenetorp, P.; and Riedel, S. 2021.
\newblock Question and Answer Test-Train Overlap in Open-Domain Question
  Answering Datasets.
\newblock In \emph{EACL}.

\bibitem[{Li et~al.(2018)Li, Duan, Zhou, Chu, Ouyang, Wang, and
  Zhou}]{li2018visual}
Li, Y.; Duan, N.; Zhou, B.; Chu, X.; Ouyang, W.; Wang, X.; and Zhou, M. 2018.
\newblock Visual Question Generation as Dual Task of Visual Question Answering.
\newblock In \emph{2018 {IEEE} Conference on Computer Vision and Pattern
  Recognition, {CVPR} 2018, Salt Lake City, UT, USA, June 18-22, 2018},
  6116--6124. {IEEE} Computer Society.

\bibitem[{Lin et~al.(2021)Lin, Ma, Lin, Yang, Pradeep, and
  Nogueira}]{Lin2021PyseriniAE}
Lin, J.; Ma, X.; Lin, S.-C.; Yang, J.-H.; Pradeep, R.; and Nogueira, R. 2021.
\newblock Pyserini: An Easy-to-Use Python Toolkit to Support Replicable IR
  Research with Sparse and Dense Representations.
\newblock \emph{ArXiv}, abs/2102.10073.

\bibitem[{Lopez et~al.(2020)Lopez, Cruz, Cruz, and
  Cheng}]{Lopez2020TransformerbasedEQ}
Lopez, L.~E.; Cruz, D.~K.; Cruz, J. C.~B.; and Cheng, C. 2020.
\newblock Transformer-based End-to-End Question Generation.
\newblock \emph{ArXiv}, abs/2005.01107.

\bibitem[{Ma et~al.(2021)Ma, Korotkov, Yang, Hall, and
  McDonald}]{Ma2021ZeroshotNP}
Ma, J.; Korotkov, I.; Yang, Y.; Hall, K.; and McDonald, R.~T. 2021.
\newblock Zero-shot Neural Passage Retrieval via Domain-targeted Synthetic
  Question Generation.
\newblock In \emph{EACL}.

\bibitem[{MacAvaney et~al.(2019)MacAvaney, Yates, Cohan, and
  Goharian}]{MacAvaney2019CEDRCE}
MacAvaney, S.; Yates, A.; Cohan, A.; and Goharian, N. 2019.
\newblock {CEDR:} Contextualized Embeddings for Document Ranking.
\newblock In Piwowarski, B.; Chevalier, M.; Gaussier, {\'{E}}.; Maarek, Y.;
  Nie, J.; and Scholer, F., eds., \emph{Proceedings of the 42nd International
  {ACM} {SIGIR} Conference on Research and Development in Information
  Retrieval, {SIGIR} 2019, Paris, France, July 21-25, 2019}, 1101--1104. {ACM}.

\bibitem[{Nogueira and Cho(2019)}]{Nogueira2019PassageRW}
Nogueira, R.; and Cho, K. 2019.
\newblock Passage Re-ranking with BERT.
\newblock \emph{ArXiv}, abs/1901.04085.

\bibitem[{Pappas, Stavropoulos, and Androutsopoulos(2020)}]{pappas2020aueb}
Pappas, D.; Stavropoulos, P.; and Androutsopoulos, I. 2020.
\newblock AUEB-NLP at BioASQ 8: Biomedical Document and Snippet Retrieval.
\newblock In \emph{CLEF (Working Notes)}.

\bibitem[{Raffel et~al.(2020)Raffel, Shazeer, Roberts, Lee, Narang, Matena,
  Zhou, Li, and Liu}]{Raffel2020ExploringTL}
Raffel, C.; Shazeer, N.~M.; Roberts, A.; Lee, K.; Narang, S.; Matena, M.; Zhou,
  Y.; Li, W.; and Liu, P.~J. 2020.
\newblock Exploring the Limits of Transfer Learning with a Unified Text-to-Text
  Transformer.
\newblock \emph{ArXiv}, abs/1910.10683.

\bibitem[{Robertson and Zaragoza(2009)}]{Robertson2009ThePR}
Robertson, S.; and Zaragoza, H. 2009.
\newblock The Probabilistic Relevance Framework: BM25 and Beyond.
\newblock \emph{Found. Trends Inf. Retr.}, 3: 333--389.

\bibitem[{Sciavolino et~al.(2021)Sciavolino, Zhong, Lee, and
  Chen}]{sciavolino2021simple}
Sciavolino, C.; Zhong, Z.; Lee, J.; and Chen, D. 2021.
\newblock Simple Entity-centric Questions Challenge Dense Retrievers.
\newblock In \emph{Empirical Methods in Natural Language Processing (EMNLP)}.

\bibitem[{Shortliffe et~al.(2014)Shortliffe, Shortliffe, Cimino, and
  Cimino}]{shortliffe2014biomedical}
Shortliffe, E.~H.; Shortliffe, E.~H.; Cimino, J.~J.; and Cimino, J.~J. 2014.
\newblock \emph{Biomedical informatics: computer applications in health care
  and biomedicine}.
\newblock Springer.

\bibitem[{Shrivastava and Li(2014)}]{Shrivastava2014AsymmetricL}
Shrivastava, A.; and Li, P. 2014.
\newblock Asymmetric {LSH} {(ALSH)} for Sublinear Time Maximum Inner Product
  Search {(MIPS)}.
\newblock In Ghahramani, Z.; Welling, M.; Cortes, C.; Lawrence, N.~D.; and
  Weinberger, K.~Q., eds., \emph{Advances in Neural Information Processing
  Systems 27: Annual Conference on Neural Information Processing Systems 2014,
  December 8-13 2014, Montreal, Quebec, Canada}, 2321--2329.

\bibitem[{Sutskever, Vinyals, and Le(2014)}]{Sutskever2014SequenceTS}
Sutskever, I.; Vinyals, O.; and Le, Q.~V. 2014.
\newblock Sequence to Sequence Learning with Neural Networks.
\newblock In Ghahramani, Z.; Welling, M.; Cortes, C.; Lawrence, N.~D.; and
  Weinberger, K.~Q., eds., \emph{Advances in Neural Information Processing
  Systems 27: Annual Conference on Neural Information Processing Systems 2014,
  December 8-13 2014, Montreal, Quebec, Canada}, 3104--3112.

\bibitem[{Thakur et~al.(2021)Thakur, Reimers, Ruckl'e, Srivastava, and
  Gurevych}]{Thakur2021BEIRAH}
Thakur, N.; Reimers, N.; Ruckl'e, A.; Srivastava, A.; and Gurevych, I. 2021.
\newblock BEIR: A Heterogenous Benchmark for Zero-shot Evaluation of
  Information Retrieval Models.
\newblock \emph{ArXiv}, abs/2104.08663.

\bibitem[{Tsatsaronis et~al.(2015)Tsatsaronis, Balikas, Malakasiotis, Partalas,
  Zschunke, Alvers, Weissenborn, Krithara, Petridis, Polychronopoulos,
  Almirantis, Pavlopoulos, Baskiotis, Gallinari, Arti{\`e}res, Ngomo, Heino,
  Gaussier, Barrio-Alvers, Schroeder, Androutsopoulos, and
  Paliouras}]{Tsatsaronis2015AnOO}
Tsatsaronis, G.; Balikas, G.; Malakasiotis, P.; Partalas, I.; Zschunke, M.;
  Alvers, M.; Weissenborn, D.; Krithara, A.; Petridis, S.; Polychronopoulos,
  D.; Almirantis, Y.; Pavlopoulos, J.; Baskiotis, N.; Gallinari, P.;
  Arti{\`e}res, T.; Ngomo, A.~N.; Heino, N.; Gaussier, {\'E}.; Barrio-Alvers,
  L.; Schroeder, M.; Androutsopoulos, I.; and Paliouras, G. 2015.
\newblock An overview of the large-scale biomedical semantic indexing and
  question answering competition.
\newblock \emph{BMC Bioinformatics}, 16.

\bibitem[{Wu et~al.(2016)Wu, Schuster, Chen, Le, Norouzi, Macherey, Krikun,
  Cao, Gao, Macherey et~al.}]{wu2016google}
Wu, Y.; Schuster, M.; Chen, Z.; Le, Q.~V.; Norouzi, M.; Macherey, W.; Krikun,
  M.; Cao, Y.; Gao, Q.; Macherey, K.; et~al. 2016.
\newblock Google's neural machine translation system: Bridging the gap between
  human and machine translation.
\newblock \emph{arXiv preprint arXiv:1609.08144}.

\bibitem[{Yang et~al.(2019)Yang, Xie, Lin, Li, Tan, Xiong, Li, and
  Lin}]{Yang2019EndtoEndOQ}
Yang, W.; Xie, Y.; Lin, A.; Li, X.; Tan, L.; Xiong, K.; Li, M.; and Lin, J.
  2019.
\newblock End-to-End Open-Domain Question Answering with {BERT}serini.
\newblock In \emph{Proceedings of the 2019 Conference of the North {A}merican
  Chapter of the Association for Computational Linguistics (Demonstrations)},
  72--77. Minneapolis, Minnesota: Association for Computational Linguistics.

\bibitem[{Zhao et~al.(2018)Zhao, Ni, Ding, and Ke}]{Zhao2018ParagraphlevelNQ}
Zhao, Y.; Ni, X.; Ding, Y.; and Ke, Q. 2018.
\newblock Paragraph-level Neural Question Generation with Maxout Pointer and
  Gated Self-attention Networks.
\newblock In \emph{Proceedings of the 2018 Conference on Empirical Methods in
  Natural Language Processing}, 3901--3910. Brussels, Belgium: Association for
  Computational Linguistics.

\end{thebibliography}
\appendix 
\section {Related Work: BioASQ and IR Systems}
BioASQ challenge~\citep{Tsatsaronis2015AnOO} is the widest challenge for Biomedical indexing and information retrieval. Since 2015, they have shared a set of questions and related documents with annotated answer passages in every year. Though plentiful IR approaches have been proposed at BioASQ, most of them are two-stage systems. The first stage is identifying initial candidates articles and the second stage is to re-rank the candidates. The re-ranking model is usually based on cross-attention model and fine-tuned for the binary classification task~\citep{nentidis2020overview}. Our focus is the first stage, i.e. large scale retrieval stage. BM25 is the widest and efficient method used in the first stage~\citep{rosso-mateus-etal-2018-mindlab,almeida2020bit,kazaryan2020transformer,pappas2020aueb}. \citet{ozyurt2020bio} also uses BM25 as the retriever, but more advanced, they iteratively retrieve relevant documents via enhanced keyword queries based LSTM network~\citep{hochreiter1997long}. 
\citet{Ma2021ZeroshotNP} use dual encoder as the neural retriever and generate large scale synthetic questions to train such a retriever. However the pure neural retriever is not outperforming BM25, hence a hybrid method is proposed which achieves the best performance in BioASQ8 challenge. Since BM25 is used as the major approach in the first stage, our main focus in this work is to compete with BM25.

\section {Statistic of Each Training Task}
Table \ref{tab:data} shows the statistic of each training task. For TempQG and AnsQG, both of them use all articles in the small corpus, but our TempQG can generate more questions than AnsQG. Because for each context, TempQG can get multiple templates (we get 10 per context) and each will be used to generate a questions. Therefore, even though using the same amount of corpus, our TempQG can generate more questions than AnsQG. 

\label{apd:dataset}
\begin{table}[h]
  \centering
  \small
  \resizebox{\linewidth}{!}{
  \begin{tabular}{lllll}
    \toprule
   & {\bf RSM} & {\bf ETM} & {\bf ICT} & {\bf BioASQ}  \\
    
    Training size & 8,54,764 & 133,084 & 133,042 & 3,243     \\
    \midrule
    & {\bf TempQG-S} & {\bf TempQG-L}  &  {\bf AnsQG-S}  &  {\bf AnsQG-L}\\
    Training size & 9,890,524 & 9,049,922 & 1,897,471 & 1,726,007  \\
    \bottomrule
  \end{tabular}
  }
  \caption{Statistic of each training task: TempQG refers to generated questions based on template-based question generation approach (Section \ref{sec:qg}), AnsQG refers to answer-awareness question generation approach (Section \ref{sec:main-result}), X-S refers to questions generated for short context and  X-L refers to long context. }
  \label{tab:data}
\end{table}

\section{Short Context Setting: What should be the context?}\label{apd:unit}
In short context setting, we use the maximum token length 128, and thus we have multiple options of context unit. 
Here, we experiment with three context units which can all be encoded by 128 tokens. {\bf Single Sentence}: we split every abstract into a single sentence. {\bf Two Sentence (w2s1)}: We apply the window-stride strategy to split every abstract into multiple pieces, where the window size is 2 sentences and stride is one sentence. For e.g., given an abstract, the first piece is the first two sentences and the next piece is the second and third sentences. Since the single sentence is the same as using the window-stride strategy with window size 1 and stride 1, we call it w1s1. {\bf Chunk}: In particular,  we use BERT tokenizer to get the tokens for an article and split them into multiple chunks where each chunk includes tokens of complete sentences and is less than 128 tokens. Every chunk is non-overlapped.  We compare two models on three different context units: DPR ( Table \ref{tab:index-unit-dpr}) and Poly-DPR ( Table \ref{tab:index-unit-polydpr}), and both model are trained on BioASQ without pre-training. For both models, w2s1 is the best context unit, which indicates that two sentences can provide more information then one sentence and more sentences are not helpful. Therefore, we use w2s1 as the context unit for short context in the main experiments presented in the paper. 

It is interesting to see that for batch 4, w1s1 is consistently better than w2s1, it might because more answers for questions in batch 4 are single sentences, and a detailed comparison among the questions in different batches will reveal more insights which will be our future work. 

Notice that we do not use a model trained on TempQG to select the best index unit since the model performance might be affected by  how the questions are generated. Our hypothesise is that since the TempQG is generated from w2s1 context, thus the performance of NRs using w2s1 might be the best. The result in Table \ref{tab:index-unit-temp}) is aligned with this hypothesise. 

\begin{table}[h]
\centering 
\small
 \resizebox{0.97\linewidth}{!}{
\begin{tabular}{@{}cllllll@{}}
    \toprule
     {\bf Unit}  & {\bf B1}   & {\bf B2} & {\bf B3} & {\bf B4} & {\bf B5} & {\bf Avg.}\\
     \toprule
        w1s1  & 55.62 & 51.0 & 53.87 & {\bf 59.32} & 49.48 & 53.86 \\
        w2s1  & {\bf 57.20} & {\bf 53.96} & {\bf 54.41} & 56.78 & {\bf 50.77} & {\bf 54.62} \\
        chunk & 53.44  &48.64 & 50.11 & 52.24 & 47.29 & 50.34 \\
    \toprule
    \end{tabular}
    }
\caption{Comparison among different context units of DPR model trained on BioASQ.}
\label{tab:index-unit-polydpr}
\end{table}

\begin{table}[h]
\centering 
\small
 \resizebox{0.97\linewidth}{!}{
\begin{tabular}{@{}cllllll@{}}
    \toprule
     {\bf Unit}  & {\bf B1}   & {\bf B2} & {\bf B3} & {\bf B4} & {\bf B5} & {\bf Avg.}\\
     \toprule
        w1s1  & 53.49   &48.65 & 52.83 & {\bf 61.46} & 47.30 & 52.75 \\
        w2s1  & {\bf 54.48} & {\bf 50.51} & {\bf 53.80} & 59.06 & {\bf 48.71} & {\bf 53.31} \\
        chunk & 50.14  & 46.72 & 51.26 & 57.57 & 46.47 & 50.43 \\
    \toprule
    \end{tabular}
    }
\caption{Comparison among different context units of Poly-DPR(K=6) model trained on BioASQ.}
\label{tab:index-unit-dpr}

\end{table}

\begin{table}[h]
\centering 
 \resizebox{0.97\linewidth}{!}{
\begin{tabular}{@{}cllllll@{}}
    \toprule
     {\bf Unit}  & {\bf B1}   & {\bf B2} & {\bf B3} & {\bf B4} & {\bf B5} & {\bf Avg.}\\
     \toprule
        w1s1  & 62.92 & 58.79 & 62.94 & {\bf 70.3} & 63.39 & 63.67 \\
        w2s1  & {\bf 63.11} & {\bf 60.35} & {\bf 62.97} & 70.2 & {\bf 63.65} & {\bf 64.06} \\
        chunk & 61.68  &59.02 & 62.25 & 70.26 & 62.21 & 63.09 \\
    \toprule
    \end{tabular}
    }
\caption{Comparison among different index units by Poly-DPR(K=6) model trained on TempQG.}
\label{tab:index-unit-temp}
\end{table}

\section{Generate Questions: How to choose the context?}\label{apd:one-que} 
Since the context is part of the input of question generator, it affects the performance of the generated questions. 
We examine two context units: a single sentence(w1s1) and two sentences(w2s2). These two contexts are obtained by the way described in Appendix \ref{apd:unit}. We are intent to use short context rather than long context (e.g. an abstract) because the question generator is trained based on short context. See Appendix \ref{apd:qg_setup} for the detail of how the generator is trained. As we observed that keeping the indexing unit in inference time as the training time yields the best performance, for a model trained on w2s1 context, we use w2s1 as the indexing unit, for a model trained on w1s1 context, we use w1s1. We trained two Poly-DPR models without any pretraining on two sets of questions, respectfully. Table \ref{tab:gq-diff-context} shows that model trained on w2s1 is much better than w1s1. We observe that the questions generator produces more duplicate questions when the context is short, as a result, the unique generated question in w2s1 is 9,890,524 while in w1s1 is 8,604,009. This might be one reason why the model trained on w2s1 is better than w1s1. Another reason might be that most snippets used to train the question generator are longer than one sentence, thus the question generator can generate better questions for w2s1 than w1s1 context. 

\begin{table}[h]
\centering 
 \resizebox{0.97\linewidth}{!}{
\begin{tabular}{@{}cllllll@{}}
    \toprule
     {\bf context}  & {\bf B1}   & {\bf B2} & {\bf B3} & {\bf B4} & {\bf B5} & {\bf Avg.}\\
     \toprule
        w1s1  & 60.65   & 57.35 & 60.95 & 64.52 & 59.41 & 60.58\\
        w2s1  & {\bf 63.11} & {\bf 60.35} & {\bf 62.97} & {\bf 70.2 }& {\bf 63.65} & {\bf 64.06} \\
        
    \toprule
    \end{tabular}
    }
\caption{Comparison between Poly-DPR(K=6) model trained on questions generated from w1s1 and w2s1.}
\label{tab:gq-diff-context}
\end{table}

\section{Main Results with Recall} \label{apd:recall}

Recall is another important metric for information retrieval, thus here, we also present the recall results of each model. As we see from Table \ref{tab:main_result_2}, in most cases, if a model has a higher MAP score than another model, then it also has a higher recall.

\begin{table*}[h]
    \centering
    \resizebox{1.0\linewidth}{!}{
    \begin{tabular}{@{}llllllllllllllll@{}}
    \toprule
    \multirow{2}{*}{\bf Size}  & \multirow{2}{*}{\bf PT} & \multirow{2}{*}{\bf FT} &  \multicolumn{2}{c}{B1 } & \multicolumn{2}{c}{B2 } & \multicolumn{2}{c}{B3 } & \multicolumn{2}{c}{B4 }& \multicolumn{2}{c}{B5 } & \multicolumn{2}{c}{Avg }\\
    \cmidrule(lr){4-5} \cmidrule(lr){6-7}  \cmidrule(lr){8-9}  \cmidrule(lr){10-11}  \cmidrule(lr){12-13} \cmidrule(lr){14-15}
    ~   & ~     & ~     &  R    &MAP    &  R    &MAP    & R     &MAP    &R  &MAP    &R      &MAP    &R  &MAP \\
    \toprule
    Small & ~     & ~     &  ~   &~    &  ~    &~    & ~     &~   &~  &~    &~      &~    &~  &~\\
    \toprule
    128 & -     & BQ    & 65.02 &54.48  & 65.0  &50.51  & 64.63 &53.8   & 70.61     &59.06  & 66.67 &48.71 & 66.39 & 53.31 \\
    128 & RSM & BQ & 72.75 & 65.94 &  {\bf 69.9} & 57.43 & 70.98 & 61.89 & 79.84 &69.01 & 71.51 &58.23 & 73.0 & 62.5  \\
    128 & -   & TempQG & 74.25 &62.92 & 69.7 &58.79 & 69.47 &62.94 & 77.62 &70.3 & 76.72 &63.39 & 73.55 &63.67  \\
    128 & RSM & TempQG & {\bf 76.01} & {\bf 64.71} & 72.48 & {\bf 64.92} & {\bf 71.52} &{\bf 64.28} & {\bf 80.45} &{\bf 73.11} & {\bf 79.51} & {\bf 66.29} & {\bf 75.99} &{\bf 66.66} \\
    \midrule
    256 & - & BQ &52.34 &35.69 & 48.16 &32.66 & 43.12 &32.26 & 52.31 &38.28 & 46.38 &30.87 & 48.46 & 33.95  \\
    
    256 & ICT & BQ & 64.92 &54.44 & 59.54 &47.37 & 60.16 &52.61 & 69.25 &53.69 & 60.85 &44.38 & 62.94 &50.50  \\
    256 & ETM & BQ & 68.95 &56.63 & 58.25 &46.63 & 63.4 &52.79 & 71.6 &56.97 & 63.19 &49.61 & 65.08 &52.53  \\
    256 & ICT+ETM & BQ &  63.78 &51.33 & 62.42 &49.43 & 57.79 &49.36 & 65.39 &53.19 & 57.19 &43.58 & 61.31 &49.38  \\
    256 & - & TempQG & 75.77 &63.95 & 70.2 &{\bf 59.51} & 71.03 &62.98 & 74.31 &66.71 & 74.45 &62.8 & 73.15 &63.19  \\
    256 & ETM & TempQG & 74.43 &64.57 & 70.75 &58.51 & {\bf 71.19} &{\bf 64.02} & {\bf 76.84} & 68.44 & 74.81 &62.6 & 73.60 &63.62  \\
    256 & ICT+ETM & TempQG & {\bf 76.24} &{\bf 64.93} & {\bf 71.53} &58.49 & {\bf 69.73} &60.18 & 76.49 & {\bf 69.42} & {\bf 76.06} &{\bf 64.87} & {\bf 74.01} & 63.58  \\
    \midrule
    BM25 & - & - & 73.82 &62.15 & 71.68 &61.3 & 72.65 &66.62 & 80.2 &74.14 & 72.86 &61.3 & 74.24 &65.10 \\
    \toprule
    Large & ~     & ~     &  ~   &~    &  ~    &~    & ~     &~   &~  &~    &~      &~    &~  &~\\
    \toprule
    128 & RSM & TempQG & {\bf 43.33} & {\bf 35.1} & 40.29 &29.07 & 41.2 &32.74 & 49.13 &33.31 & 46.04 &35.54 & 44.0 &33.15  \\
    BM25 & - & - & 38.94 &28.5 & 44.31 &27.82 & {\bf 48.45} &37.97 & {\bf 54.65} &41.91 & 49.02 &35.42 & 47.07 &34.32 \\
    Hybrid & ~ & - & 40.59 &30.02 & {\bf 47.92} & {\bf 31.31} & 46.55 & {\bf 39.79} & 54.02 &{\bf 42.18} & {\bf 49.37} & {\bf 37.99} & {\bf 47.69} & {\bf 36.26} \\
    \toprule
    \end{tabular}
    }
    \caption{Performance of model trained by different techniques on BioASQ8 testing set, CS: context setting, PT: pretraining task, FT: finetuning task, Bi: batch i. BQ: BioASQ}
    \label{tab:main_result_2}
\end{table*}

\section{Experiment Setup} \label{apd:setup}

\subsection{Neural Retriever Training}
All neural retrievers were trained at Quadro RTX 8000 GPUs. For all pretraining or fine-tuning tasks, we set the learning rate (lr) to be 2e-05,  max gradient norm to be 2.0. For pretraining tasks, we set the epoch to be 20, and select the best model based on the evaluation of the training set of BioASQ. When we train (or fine-tune) models on TempQG, we set the epoch to be 1, and save a model per 1/5 epoch, and the best model is selected based on the evaluation on the training set of BioASQ (for most experiments, the last model is the best). We found that training with more epoch does not increase model performance. When we train (or fine-tune) models on BioASQ, we set the epoch to be 20 and use the model of the last epoch for evaluation.  Following \citep{Karpukhin2020DensePR}, we use the in-batch negative training strategy and they find that larger batch size yields better performance. Thus we use the largest batch size as allowed by our machine, i.e. batch size is 256 for long context and 768 for short context.  

\subsection{Template based Question Generation Training}\label{apd:qg_setup}

As described in Section \ref{sec:qg}, our template-based question generation consists of two components, a template selection model and a question generation model. We describe each as bellow. 

\paragraph{Template Selection Model }We use a Poly-DPR as the template section model and the training data is from BioASQ. Specifically, in BioASQ, each question has a list of gold snippets and given a  question we first derive a template in the way described in Section \ref{sec:qg}, then we take each snippet as query, and the template as the context. In total, we have 41,915 query-template training pairs and the size of unique templates is 1,052. 

We train the template selection model at Quadro RTX 8000 GPUs, and we set lr as 2e-05,  max gradient norm as 2.0, epoch as 20, bs as 20, the maximum length of encoding as 128. 

\paragraph{Question Generation Model} The generator is based on  T5-small ~\citep{Raffel2020ExploringTL}. For each context, we first use the template section model to obtain $N$ templates, then we concatenate each template with the question independently such that the input to the model is  $\{``template": template, ``context": text\}$. 
We train the model at GTX1080 and V100 NVIDIA GPUs and set lr as 1e-4, gradient accumulation steps as 8, bs as 32, epoch as 5. 

Using a larger T5 model, like T5-base or T5-large might potentially generate better questions, we leave this as future work. 

Notice that for AnsQG, we directly use the public available implementation\footnote{https://github.com/patil-suraj/question\_generation} to generate questions for PubMed articles. To have fair comparison, we use the same PubMed articles to generate questions as TempQG. 
\section{Other Analysis}  \label{apd:analysis}



\subsection{Training with Non-Overlapped Data}

\citet{Lewis2021QuestionAA} find that DPR performs much worse on the set of train-test non-overlapped questions than overlapped ones.  
Here, we assume a scenario where the training corpus has no overlap with the testing corpus. As described in Section \ref{sec:exp} that the small testing corpus has 133,084 articles, we randomly select other 133,084 articles which does not include any articles in the testing corpus, We call this set of new articles a non-overlapped corpus. Then we generate questions for these new articles by our template-based question generator, which ends up with 9,695,155 questions.  We train a Poly-DPR on the new questions and evaluate it on the testing corpus. Table \ref{tab:non-overlapped} shows the comparison of the model trained on a non-overlapped corpus with the model trained on the testing corpus. Unsurprisingly, the model trained on testing corpus is better, yet the gap between these two models is not significant, which indicates the effectiveness of our generated questions and the Poly-DPR architecture. 

\begin{table}[h]
\centering 
 \resizebox{0.97\linewidth}{!}{
\begin{tabular}{@{}cllllll@{}}
    \toprule
     {\bf corpus}  & {\bf B1}   & {\bf B2} & {\bf B3} & {\bf B4} & {\bf B5} & {\bf Avg.}\\
     \toprule
        non-overlapped corpus & 63.08   & {\bf 61.41} & 60.33 & 68.18 & 61.42 & 62.88\\
        testing corpus  &  {\bf 63.11} &  60.35& {\bf 62.97} & {\bf 70.2 }& {\bf 63.65} & {\bf 64.06} \\
        
    \toprule
    \end{tabular}
    }
\caption{Comparison between Poly-DPR(K=6) model trained on questions generated from non-overlapped corpus and testing corpus.}
\label{tab:non-overlapped}
\end{table}

\subsection{Sequential Training or Multi-Task Training} 

Since our generated questions are large, even larger than the data of pretraining tasks, this allows us to train a model in two different ways. {\bf Sequence Training}: we train a model on the pretraining task data and then on the generated question data. In such a way, we expect the model to learn the general knowledge from the pre-training task then learn the knowledge for the end task. {\bf Multi-Task Training}: we train the model on pretraining tasks and end-task simultaneously.  We compare two training strategies on short and long context settings. Table \ref{tab:sequence-multi} shows the comparison where we see that the sequential training strategy is better than the multi-task training strategy on average. The difference between the two training strategies is larger in short context settings than in long context. The reason might be that in the long context, the data from the end-task is much larger than the pretraining task (see Table \ref{tab:data}) which dominates the model performance, thus two strategies do not make difference. 

\begin{table}[ht!]
\centering
 \resizebox{0.97\linewidth}{!}{
\begin{tabular}{@{}llllllll@{}}
    \toprule
     {\bf Size}  & {\bf Strategy} & {\bf B1}   & {\bf B2} & {\bf B3} & {\bf B4} & {\bf B5} & {\bf Avg.}\\
     \toprule
        128   & Sequential   & {\bf 64.71}   & {\bf 64.92} & {\bf 64.28} & {\bf 73.11} & 62.29 & {\bf 66.66} \\
        128   & Multi-Task   & 64.19 & 59.08 & 63.88 & 69.43 & {\bf 64.33} & 64.18\\
        
        \midrule
        256   & Sequential  &  64.57 & {\bf 58.51} & {\bf 64.02} & 68.44 &  62.6 & {\bf 63.62} \\
        256   & Multi-Task & {\bf 64.93} & 58.49 & 60.18 &{\bf  69.42} & {\bf 64.87} & 63.58\\
        
    \toprule
    \end{tabular}
    }
\caption{Comparison of two training strategies on short and long context.}
\label{tab:sequence-multi}
\end{table}

\subsection{Final Prediction of Document Retrieval}
In case when we split each document into multiple indexing such as w1s1, w2s1, chunk-128 and chunk-256, it might happen that multiple indexing from one document will be retrieved. If that happens, we choose the highest score as the document score and output the top 10 documents as the final predictions. BioASQ require maximum 10 documents per question. 

\section{More Examples for Question Generation}

Here we provides more examples of our template-based question generation to show that our approach can generate more diverse and biomedical domain oriented questions.  

\begin{table*}[h]
    \centering 
    \small
    \resizebox{0.97\linewidth}{!}{
    \begin{tabular}{@{}p{0.01\linewidth}p{0.68\linewidth}p{0.13\linewidth}p{0.25\linewidth}@{}}
        \toprule 
        \textbf{\#} &\textbf{Context} & \textbf{Template} & \textbf{Generated Question} \\
        \toprule
        1 & Alpers syndrome is an autosomal recessive mitochondrial DNA depletion disorder that affects children and young adults. It is characterized by a progressive, fatal brain and liver disease.
            & what is the inheritance of the \_ deficiency   
            & What is the inheritance of the Alpers syndrome? \\
        \midrule
        1 & Thus, curcumin enhanced the efficacy of anti-tumor drugs imatinib-mesylate and cytarabine through the inhibition of the tumor markers. 
            & could \_ be used in \_ against \_  
            & Could curcumin be used in anti-tumor drugs against cytarabine? \\
        \midrule
        1 & The lysosomal-membrane protein type 2A (LAMP-2A) acts as the receptor for the substrates of chaperone-mediated autophagy (CMA), which should undergo unfolding before crossing the lysosomal membrane and reaching the lumen for degradation.
            & which receptor is targeted by \_ 
            & Which receptor is targeted by LAMP-2A? \\
        \midrule
        2 & Is Tokuhashi score suitable for evaluation of life expectancy before surgery in Iranian patients with spinal metastases? One of the most important selection criteria for spinal metastases surgery is life expectancy and the most important system for this prediction has been proposed by Tokuhashi.
            & what is evaluated with \_ 
            & What is the Tokuhashi score? \\
        \midrule
        
        3 & MiR-208 is involved in the shift toward a fetal gene expression pattern in contractile proteins in heart failure. MiR-1 influences susceptibility to cardiac arrhythmias after myocardial infarction.
             & is \_ influenced by \_ 
             & Is myocardial infarction influenced by MiR-1? \\
        \midrule
        4 & Future research needs to examine the mechanisms by which these therapies reduce incontinence and whether combining behavioral and drug treatment will result in better outcomes than either therapy alone.
            & can \_ alleviate \_ 
            & Can drug therapy alleviate incontinence? \\
        \midrule
        3 & Lambert-Eaton myasthenic syndrome (LEMS) is a pre-synaptic disorder of the neuromuscular and autonomic transmission mediated by antibodies to voltage-gated calcium channels at the motor nerve terminal. 
        LEMS is a quite rare and probably under-diagnosed disease: the onset may be slow and clinical signs are typically fluctuating, thus adding to the delay in diagnosis.
            & \_ is diagnosed in which \_ 
            & Lambert-Eaton myasthenic syndrome is diagnosed in which neuromuscular and autonomic pathways? \\
        \midrule
        6 & Analysis of DSB sites in three regions of the yeast genome indicated that breaks occur at or near many potential transcription promoters and that DSBs initiate most, if not all, meiotic recombination. DSB sites displayed deoxyribonuclease I hypersensitivity in chromatin from mitotic and meiotic cells, and changes in chromatin structure produced parallel changes in the occurrence of DSBs.
            & does \_ dna replication stress inhibit \_ 
            & Does deoxyribonuclease I hypersensitivity inhibit chromatin recombination? \\
        \bottomrule
    \end{tabular}
    }
    \caption{Ilustrative examples for templates and questions generated by TempQG.
    }
    \label{tab:ques-example-more}
\end{table*}

\section{More Examples of BM25 and NR failures}

Here we provide more examples of BM25 and NR failures. We find that the type of questions where BM25 and NR finds difficult are quite different, and thus the simple hybrid method that we proposed can take advantage from both. 

\begin{table*}[t]
    \centering 
    \small
    \resizebox{\linewidth}{!}{
    \begin{tabular}{@{}p{0.11\linewidth}p{0.24\linewidth}p{0.64\linewidth}@{}}
        \toprule 
        \textbf{Category} & \textbf{Question} & \textbf{Explanation} \\
        \toprule
        
        BM25-1 &  What is minodixil approved for?& minodixil is a typo, the correct one is minoxidil\\
        BM25-2 & List five proteins with antioxidant properties?  & BM25 fails to connect proteins and antioxidant properties, and retrieves documents all related to antioxidant, however, they are not about proteins nor antioxidant proteins.\\
        BM25-2 & Has ZP-PTH been tested in a phase II clinical trial? & BM25 retrieves documents about phase II clinical trial but not about  ZP-PTH.\\
        BM25-3 & How  large  is  a  lncRNAs? & BM25 retrieves document about lncRNAs but not about how large it is. \\
        BM25-3 & Is indinavir effective for treatment of amyotrophic lateral sclerosis? & BM25 retrieves documents related to amyotrophic lateral sclerosis but not the treatments of amyotrophic lateral sclerosis. \\
        BM25-3 & What virus is the  Gardisil vaccine used for? &  BM25 retrieves documents about virus, but not Gardisil vaccine neither documents are about both of them.\\
        PolyDPR-1 & What is Xanamem? & NR fails to retrieve any document related to Xanamem, rather, it retrieves documents that lexical similar to Xanamem such as Ximenia, Xadago, and Xenopus.\\
        PolyDPR-1 & What is Soluvia? & NR fails to retrieve any document related to Soluvia, rather, it retrieves documents that lexical similar to Soluvia such as Solexa, Saliva, and Viagra.\\
        PolyDPR-2 & Does an interferon (IFN) signature exist for SLE patients? & NR ranks documents about interferon higher than documents of SLE patients and documents of both. In the retrieved documents, interferon appears rather frequently.\\
        \bottomrule
    \end{tabular}
    }
    \caption{Failure cases and explanations for BM25 and Poly-DPR for various categories. More examples in Appendix.
    }
    \label{tab:error_analysis}
\end{table*}

\end{document}